\documentclass[letterpaper,aps,prc,amsmath,superscriptaddress,floatfix]{revtex4}

\usepackage{graphicx}   
\usepackage{multirow}
\usepackage{bm}
\usepackage{subfigure}  
\usepackage{color}
\usepackage{epstopdf}   
\usepackage{amsmath}    
\usepackage{amsfonts}   
\usepackage{amssymb}    
\usepackage{amscd}      
\usepackage{natbib}     
\usepackage{longtable}  
\usepackage{floatrow}   
\usepackage{hyperref}   

\newcommand{\figwidth}{0.85\columnwidth}

\newcommand{\reaction}[1]{$^{15}$N(p,$\gamma_0$)$^{16}$O}

\begin{document}

\title{Constraining the S factor of $^{15}$N(p,$\gamma$)$^{16}$O at
Astrophysical Energies}

\author{P.~J. LeBlanc}
\email[email at: ]{leblanc.pj@gmail.com}
\affiliation{University of Notre Dame, Department of Physics,
Notre Dame, IN 46556, USA}

\author{G. Imbriani}
\affiliation{University of Notre Dame, Department of Physics,
Notre Dame, IN 46556, USA}
\affiliation{Universit\`a degli Studi di Napoli ``Frederico II", and INFN, Napoli, Italy}

\author{J. G\"{o}rres}
\affiliation{University of Notre Dame, Department of Physics,
Notre Dame, IN 46556, USA}

\author{M. Junker}
\affiliation{INFN, Laboratori Nazionali del Gran Sasso (LNGS), Assergi (AQ), Italy}

\author{R. Azuma}
\affiliation{University of Notre Dame, Department of Physics,
Notre Dame, IN 46556, USA}
\affiliation{Department of Physics, University of Toronto, Toronto, Ontario M55
1A7, Canada}

\author{M. Beard}
\affiliation{University of Notre Dame, Department of Physics,
Notre Dame, IN 46556, USA}

\author{D. Bemmerer}
\affiliation{Forschungszentrum Dresden-Rossendorf, Dresden, Germany}

\author{A. Best}
\affiliation{University of Notre Dame, Department of Physics,
Notre Dame, IN 46556, USA}

\author{C. Broggini}
\affiliation{INFN, Padova, Italy}

\author{A. Caciolli}
\affiliation{Universit\'a degli Studi di Padova and INFN,  Padova, Italy}

\author{P. Corvisiero}
\author{H. Costantini}
\affiliation{Universit\'a degli Studi di Genova and INFN,  Genova, Italy}

\author{M. Couder}
\author{R. deBoer}
\affiliation{University of Notre Dame, Department of Physics,
Notre Dame, IN 46556, USA}

\author{Z. Elekes}
\affiliation{Institute of Nuclear Research (ATOMKI), Debrecen, Hungary}

\author{S. Falahat}
\affiliation{University of Notre Dame, Department of Physics,
Notre Dame, IN 46556, USA}
\affiliation{Max-Planck-Institut f\"ur Chemie, Mainz, Germany}

\author{A. Formicola}
\affiliation{INFN, Laboratori Nazionali del Gran Sasso (LNGS), Assergi (AQ), Italy}

\author{Zs. F\"ul\"op}
\affiliation{Institute of Nuclear Research (ATOMKI), Debrecen, Hungary}

\author{G. Gervino}
\affiliation{Universit\'a degli Studi di Torino and INFN,  Torino, Italy}

\author{A. Guglielmetti}
\affiliation{Universit\`a degli Studi di Milano and INFN, Sez. di Milano,
Milan, Italy}

\author{C. Gustavino}
\affiliation{INFN, Laboratori Nazionali del Gran Sasso (LNGS), Assergi (AQ), Italy}

\author{Gy. Gy\"urky}
\affiliation{Institute of Nuclear Research (ATOMKI), Debrecen, Hungary}

\author{F. K\"{a}ppeler}
\affiliation{Forschungszentrum Karlsruhe, Institut f\"{u}r
Kernphysik, Karlsruhe, Germany}

\author{A. Kontos}
\affiliation{University of Notre Dame, Department of Physics, Notre Dame, IN 46556, USA}

\author{R. Kuntz}
\affiliation{Institute f\"ur Experimentalphysik III, Ruhr-Universit\"at Bochum, Bochum, Germany}

\author{H. Leiste}
\affiliation{Forschungszentrum Karlsruhe, Institut f\"{u}r
Materialforschung I, Karlsruhe, Germany}

\author{A. Lemut}
\affiliation{Universit\'a degli Studi di Genova and INFN, Genova, Italy}

\author{Q. Li}
\affiliation{University of Notre Dame, Department of Physics, Notre Dame, IN 46556, USA}

\author{B. Limata}
\affiliation{Universit\`a degli Studi di Napoli ``Frederico II", and INFN, Napoli, Italy}

\author{M. Marta}
\affiliation{Forschungszentrum Dresden-Rossendorf, Dresden, Germany}

\author{C. Mazzocchi}
\affiliation{Universit\`a degli Studi di Milano and INFN, Sez. di Milano,
Milan, Italy}

\author{R. Menegazzo}
\affiliation{INFN, Padova, Italy}

\author{S. O'Brien}
\author{A. Palumbo}
\affiliation{University of Notre Dame, Department of Physics,
Notre Dame, IN 46556, USA}

\author{P. Prati}
\affiliation{Universit\`a degli Studi di Genova and INFN, Genova, Italy}

\author{V. Roca}
\affiliation{Universit\`a degli Studi di Napoli ``Frederico II", and INFN, Napoli, Italy}

\author{C. Rolfs}
\affiliation{Institute f\"ur Experimentalphysik III, Ruhr-Universit\"at Bochum, Bochum, Germany}

\author{C. Rossi Alvarez}
\affiliation{INFN, Padova, Italy}

\author{E. Somorjai}
\affiliation{Institute of Nuclear Research (ATOMKI), Debrecen, Hungary}

\author{E. Stech}
\affiliation{University of Notre Dame, Department of Physics,
Notre Dame, IN 46556, USA}

\author{O. Straniero}
\affiliation{Osservatorio Astronomico di Collurania. Teramo, and INFN Napoli, Italy}

\author{F. Strieder}
\affiliation{Institute f\"ur Experimentalphysik III, Ruhr-Universit\"at Bochum, Bochum, Germany}

\author{W. Tan}
\affiliation{University of Notre Dame, Department of Physics,
Notre Dame, IN 46556, USA}

\author{F. Terrasi}
\affiliation{Seconda Universit\`a di Napoli, Caserta and INFN, Napoli, Italy}

\author{H.P. Trautvetter}
\affiliation{Institute f\"ur Experimentalphysik III, Ruhr-Universit\"at Bochum, Bochum, Germany}

\author{E. Uberseder}
\affiliation{University of Notre Dame, Department of Physics,
Notre Dame, IN 46556, USA}

\author{M. Wiescher}
\affiliation{University of Notre Dame, Department of Physics,
Notre Dame, IN 46556, USA}

\date{\today}

\begin{abstract}
{\footnotesize The $^{15}$N(p,$\gamma$)$^{16}$O reaction
represents a break out reaction linking the first and second cycle
of the CNO cycles redistributing the carbon and nitrogen
abundances into the oxygen range. The reaction is dominated by two
broad resonances at $E_p$ = 338 keV and 1028 keV and a Direct
Capture contribution to the ground state of $^{16}$O. Interference
effects between these contributions in both the low energy region
(E$_p$ $<$ 338 keV) and in between the two resonances (338 $<$
E$_p$ $<$ 1028 keV) can dramatically effect the extrapolation to
energies of astrophysical interest. To facilitate a reliable
extrapolation the $^{15}$N(p,$\gamma$)$^{16}$O reaction has been
remeasured covering the energy range from E$_p$=1800 keV down to
130 keV. The results have been analyzed in the framework of a
multi-level R-matrix theory and a S(0) value of 39.6 keV b has
been found.}
\end{abstract}

\maketitle

\section{\label{sec:intro}\bfseries{Introduction}}
The energy production and nucleosynthesis in stars is
characterized by nuclear reaction sequences which determine
the subsequent phases of stellar evolution. Energy production
during the first hydrogen burning phase takes place through the
fusion of four protons into helium. This occurs either through
the pp-chains or the CNO cycles. The pp-chains dominate hydrogen
burning in first generation stars with primordial abundance
distributions and in low mass, M$\le$1.5 M$_{\odot}$, stars. The CNO
cycles dominate the energy production in more massive, M$\ge$1.5
M$_{\odot}$, second or later generation stars with an appreciable
abundance of CNO isotopes. The CNO cycles are characterized by
sequences of radiative capture reactions and $\beta$-decay
processes as shown in Figure~\cite{Cauldrons88}.

\begin{figure}[tbh]
\includegraphics[scale=0.35]{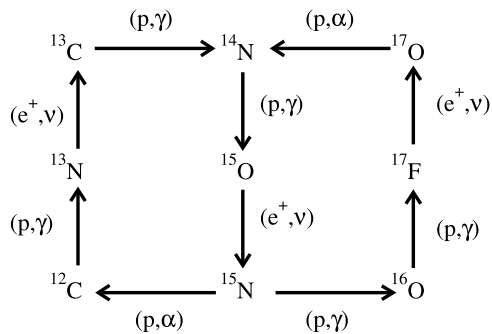}
\caption{Diagram of the CNO bi cycle~\cite{Cauldrons88}}
\label{fig:CNO_diagram}
\end{figure}

At stellar temperatures the $^{14}$N(p,$\gamma$)$^{15}$O reaction
is the slowest process in the cycle, defining the time scale and
the overall energy production rate~\cite{Imbriani04,
Imbriani05,Runkle05}. This reaction is therefore of importance for
the interpretation of CNO burning. The proton capture by $^{15}$N
is relevant since it is a branch point linking the first CNO or CN
cycle with the second CNO, or NO cycle as shown in
Figure~\cite{Cauldrons88}~\cite{CaF62}. This branching has always
been a matter of debate since both reactions are characterized by
strong low energy resonances.

The reaction rate of $^{15}$N(p,$\alpha$)$^{12}$C is determined by
two broad low energy s-wave resonances at E$_p$ = 338 keV and 1028
keV, populating the J$^{\pi}$ = 1$^-$ states at 12.44 and 13.09
MeV, respectively, in the $^{16}$O compound nucleus. There have
been a number of low energy measurements~\cite{SFL52},
~\cite{Zys79}, and~\cite{RBL82} which provide the basis of the
present rate in the literature~\cite{CaF88,Angulo99}. Recently,
three lower energy points were derived from an indirect ``Trojan
Horse Method (THM)"~\cite{LaCognata09} which are consistent with
low energy data~\cite{SFL52,Zys79,RBL82}.

The competing $^{15}$N(p,$\gamma$)$^{16}$O reaction decays
predominately to the ground state of $^{16}$O and exhibits the
same two resonances but in addition is expected to have a strong
non-resonant direct capture component~\cite{Rolfs74}.  The presently
available low energy cross section data from~\cite{Hebbard60},~\cite{Brochard73},
and~\cite{Rolfs74} differ substantially at
lower energies. This poses difficulties for a reliable
extrapolation of the cross section towards the stellar energy
range. An extrapolation was performed~\cite{Rolfs74} using a two
level Breit-Wigner formalism taking into account the direct capture
contribution. The present reaction rate for
$^{15}$N(p,$\gamma$)$^{16}$O ~\cite{Angulo99} relies entirely on
the predictions of~\cite{Rolfs74}.

To reduce the uncertainty in the strength of the direct capture
term single-particle transfer reactions have been performed
~\cite{MBB08} to determine the proton Asymptotic Normalization
Coefficient (ANC) for the ground state of $^{16}$O.  With this ANC
value $C_{p_{1/2}}^2 = (192 \pm 26)$ fm$^{-1}$, R-matrix fits to
the $^{15}$N(p,$\gamma$)$^{16}$O data have been
performed~\cite{MBB08} which resulted in smaller values for the
low energy cross section than those obtained by~\cite{Rolfs74}.
This result was confirmed by an independent R-matrix analysis of
the existing data~\cite{Barker08b}.  This conclusion was
furthermore supported by a study in which both the (p,$\alpha$)
and (p,$\gamma$) reactions were analyzed simultaneously with a
multi-level, multi-channel R-matrix formalism~\cite{Simpson06}.

Low energy data points were extracted from a
re-analysis~\cite{Bemmerer09} of a study of
$^{14}$N(p,$\gamma$)$^{15}$O performed at the LUNA underground
accelerator facility at the Gran Sasso laboratory.  This
experiment was performed using a windowless differentially pumped
gas target with natural nitrogen gas.  For detecting the $\gamma$-ray
signal, a large BGO scintillator detector was used to observe the
characteristic $\gamma$ decay in summing mode~\cite{Keinonen76}.
Since the natural abundance of $^{15}$N is low, the yield of the
$^{15}$N(p,$\gamma$) signal was weak and overshadowed by beam
induced background yield from proton capture reactions on target
impurities and from the $^{14}$N(p,$\gamma$) reaction. The
measurements were therefore limited to energies below 230 keV. The
proposed cross section results are clearly lower than the values
obtained by~\cite{Rolfs74} but also slightly lower than the results
of~\cite{Hebbard60} and~\cite{Brochard73}. Given the strong
background conditions, the results could not be normalized to the
known on-resonance yield at 338 keV and systematic errors in the
data cannot be excluded.

Because of the inconsistencies in the existing data and the
uncertainties of an R-matrix analysis based on these existing
data, we have performed a new study of the
$^{15}$N(p,$\gamma$)$^{16}$O over a wide energy range using high
resolution Ge detectors. In the following section the experimental
approach at the Notre Dame Nuclear Science Laboratory (NSL) and
the LUNA II facility at the Gran Sasso underground laboratory will
be discussed. This will be followed by a discussion of the
experimental data. In the last section the stellar nuclear reaction
rate based on the present data will be calculated and compared
with existing results.

\section{\label{sec:exp}\bfseries{Experimental Setup}}

\subsection{\bfseries{Accelerators and Experimental Setup}}

The experiment was performed at two separate facilities. At the
University of Notre Dame the 4 MV KN Van de Graaff accelerator
provided proton beams in an energy range of 700 to 1800 keV with
beam intensities limited to $\leq$10 $\mu$A on target because of
the high count rate in the Ge detector from the
$^{15}$N(p,$\alpha_1\gamma$)$^{12}$C reaction. The energy
calibration of this machine was established to better than 1 keV
using the well known $^{27}$Al(p,$\gamma$)$^{28}$Si resonance at
992 keV~\cite{Keinonen76}. The 1 MV JN Van de Graaff accelerator
at Notre Dame was used in the range of 285 keV to 700 keV with
protons beams of 20 $\mu$A. The energy of the this machine was
calibrated using the well known
$^{15}$N(p,$\alpha_1\gamma$)$^{12}$C resonance at 429
keV~\cite{Ajzenberg86}.

The LUNA II facility~\cite{Costantini09}, located in the Gran
Sasso National Laboratory, uses a high current 400 kV, Cockroft-
Walton type accelerator. The accelerator provided proton beam
currents on target of up to 200 $\mu$A in the energy range of 130
to 400 keV.  In addition to the high beam output, the accelerator
is extremely stable, and the voltage is known with an accuracy of
about 300 eV.

The experimental setup in both experiments was very similar. The
targets were water cooled and mounted at 45$^{\circ}$ with respect
to the beam direction. At Notre Dame the position of the beam on
the target was defined by a set of vertical and horizontal slits.
The beam was swept horizontally and vertically across a target
area of 1 cm$^2$ by steerers in order to dissipate power over a
large target area.  At LUNA the ion beam optics provided a
de-focused beam on target and no beam sweeping was applied. To
avoid the build-up of impurities on the target a Cu finger, cooled
to LN$_2$ temperatures, was placed along the inside of the
beamline extending as close to the target as possible.  In
addition, a bias voltage of about -400 V was applied to the
isolated cold finger to suppress the secondary electrons ejected
from the target due to proton bombardment.

\subsection{\bfseries{Targets}}

The Ti$^{15}$N targets were fabricated at the Forschungszentrum
Karlsruhe by reactive sputtering of Ti in a Nitrogen atmosphere
enriched in $^{15}$N to 99.95\%. The stoichiometry was analyzed
using Auger electron spectroscopy to confirm the composition. This
test was performed on two target spots, one which had been exposed
to beam and one which was not exposed. The results agreed within
$\le$2$\%$ with the nominal stoichiometry of 1:1. Isotopic
abundances were experimentally verified by comparing the yield of
the $^{14}$N(p,$\gamma)^{15}$O, 278 keV
resonance~\cite{Imbriani05} from the enriched targets with that
obtained using a target produced with a natural nitrogen gas. The
results of this measurement showed an abundance of $\le$ 2\% of
$^{14}$N for the thin targets corresponding to a $^{15}$N
enrichment of $\ge$ 98\% in agreement with the quote of the
supplier. For the thick target used at LUNA, $^{14}$N and $^{15}$N
enrichment of 17.4 $\pm$ 2.0\% and 82.6 $\pm$ 2.0\% were found,
respectively, most likely caused by a contamination of the
enriched gas during sputtering.

\begin{figure}
\centering
\mbox{\subfigure{\includegraphics[width=0.45\columnwidth]{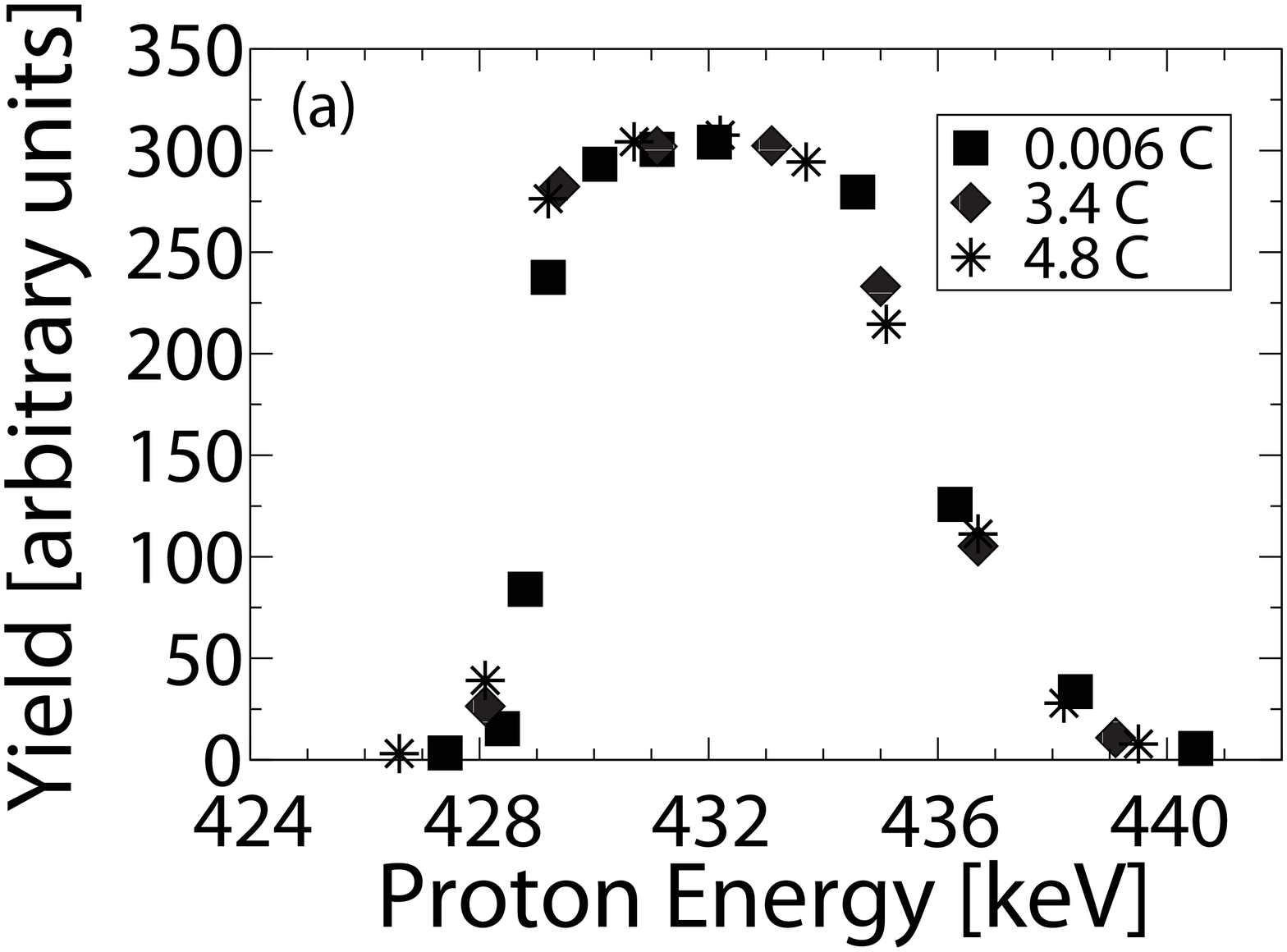}}\quad
\subfigure{\includegraphics[width=0.45\columnwidth]{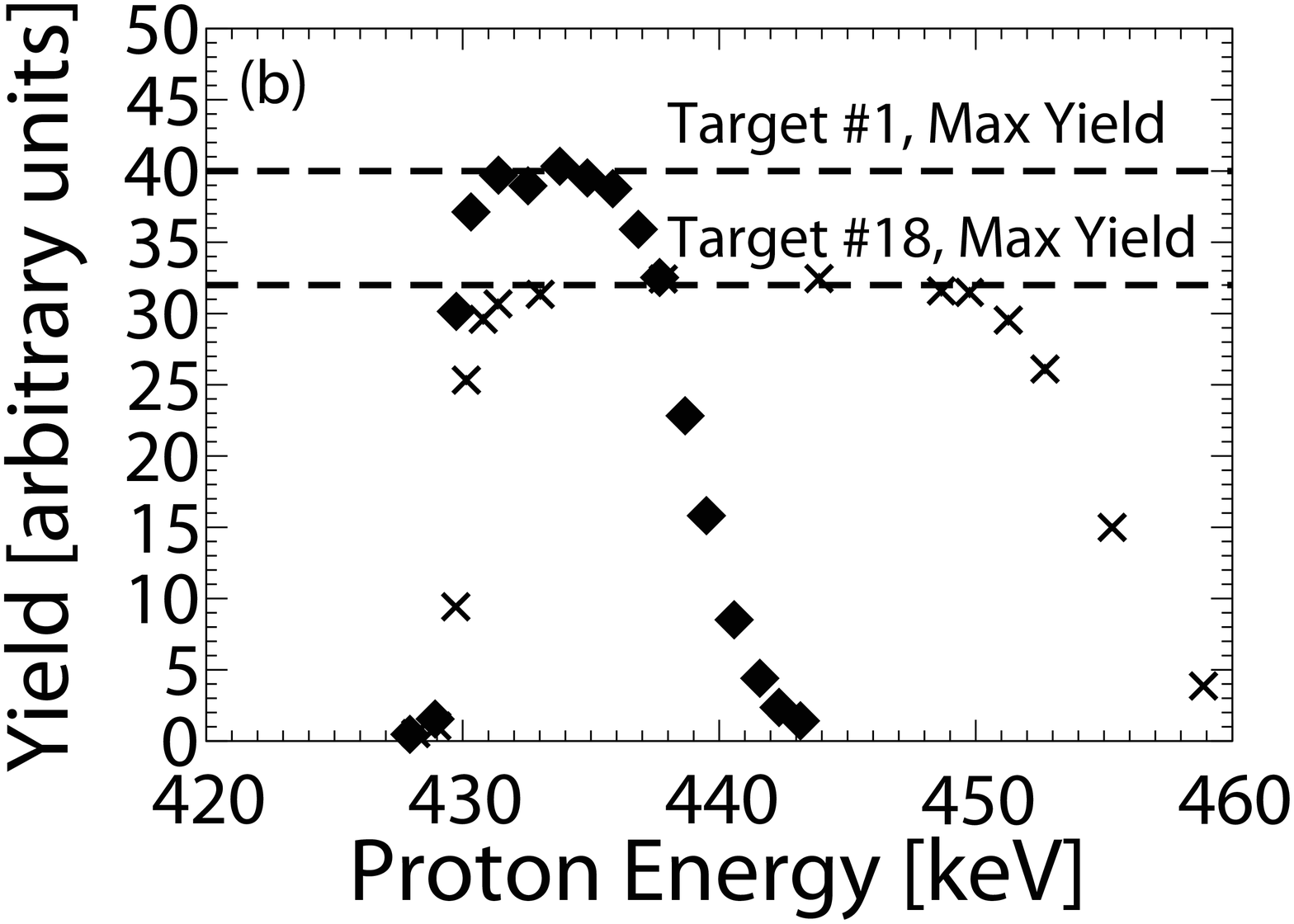} }}
\caption{Notre Dame target (left) has a measured width of approximately 7.2 keV
at E$_p$ = 430 keV.  Squares, diamonds, and stars correspond to 0.006 C, 3.4 C,
and 4.8 C of accumulated charge on the target, respectively.  LUNA thin (right,
diamonds) and thick (right, crosses) targets have measured widths of
approximately 9.5 keV, and approximately 25 keV, respectively.  The difference
in stoichiometry between the two targets can be seen by comparing the plateau
heights from the same resonance scan of each target (right).}
\label{fig:430keV_scans}
\end{figure}

\begin{figure}[tbp]
\includegraphics[width=\figwidth]{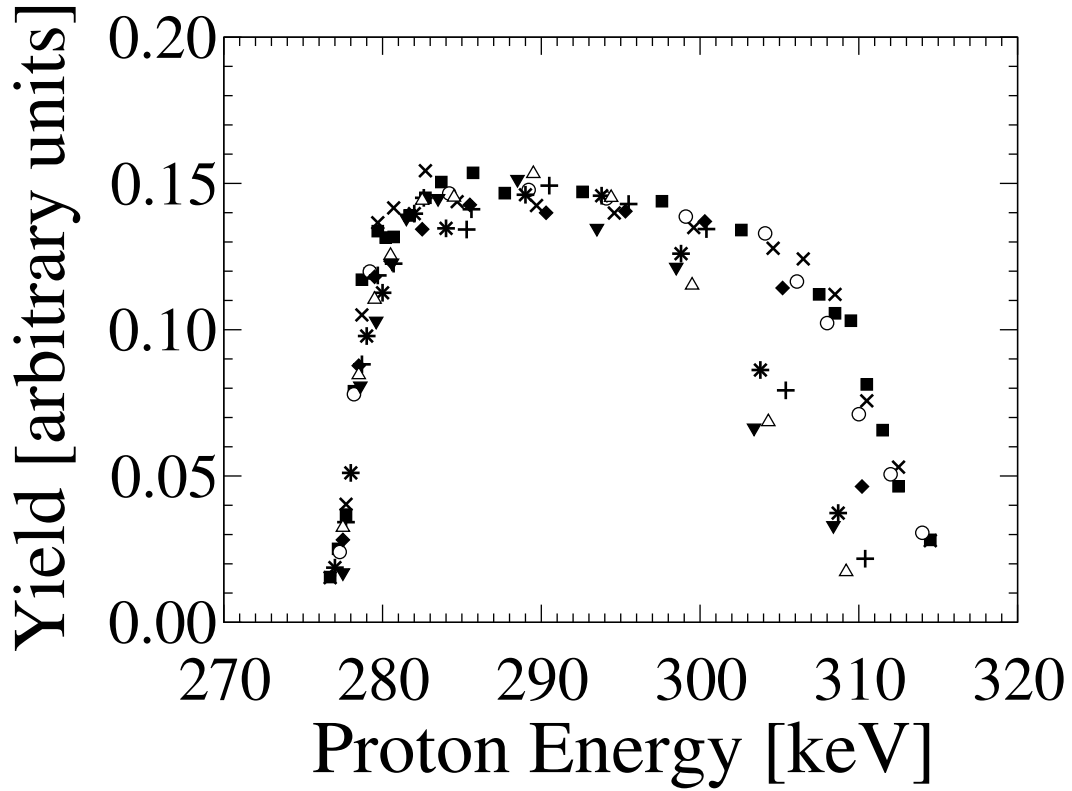}
\caption{Targets scans of thick target (18) using the
$^{14}$N(p,$\gamma)^{15}$O, E$_p$ = 278 keV resonance.  The target
was stable until about 20 C, and went through significant
deterioration. Squares represent the initial target scan, while
down triangles represent the scan after 65 C on target \label{fig:luna_target_scans}}
\end{figure}

The thicknesses of all TiN targets were measured using the narrow
$^{15}$N(p,$\alpha_1\gamma$)$^{12}$C resonance at 429
keV~\cite{Ajzenberg86}. The target used at Notre Dame had a
thickness of 7.2 $\pm$ 0.3 keV at E$_p$ = 429 keV and the two
targets used at LUNA had thicknesses of 9.5 $\pm$ 0.4 keV and 24.8
$\pm$ 0.5 keV at 429 keV, respectively. The stability of the Notre
Dame targets was checked continuously during the course of the
experiment using the $^{15}$N(p,$\alpha_1\gamma$)$^{12}$C
resonance.  The thin LUNA target was monitored by rescanning the
top of the 338 keV resonance in $^{15}$N(p,$\gamma$)$^{16}$O.  The
thick LUNA target could also be monitored using the
$^{14}$N(p,$\gamma$)$^{15}$O resonance at E$_p$ = 278 keV because
of the large $^{14}$N content (17\%) of this target.  Because of
the relatively low power density delivered at Notre Dame, the
target saw virtually no degradation over the experiment with an
accumulated charge of 5 C (see Figure~\ref{fig:430keV_scans}) and
no yield corrections were necessary. During the LUNA experiment
with significantly higher beam currents, the thickness of the thin
target was reduced by 27$\%$ after an accumulated charge of 17 C
and that of the thick target by 30$\%$ after an accumulated charge
of 65 C (see Figure~\ref{fig:luna_target_scans}).

\subsection{\bfseries{Detectors}}
\begin{figure}[tbp]
\includegraphics[width=0.70\columnwidth]{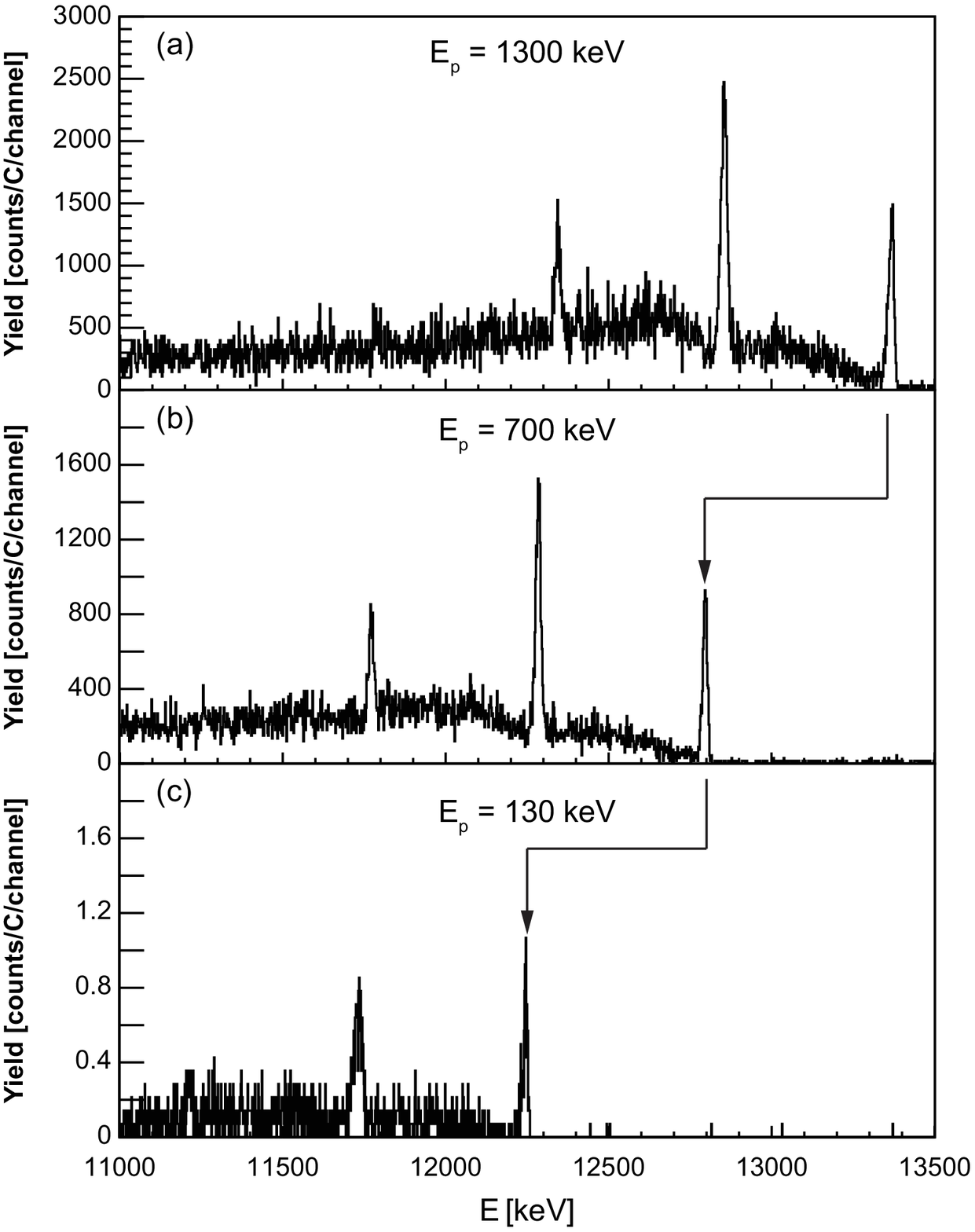}
\caption{Several sample spectra from the present
$^{15}$N(p,$\gamma$)$^{16}$O experiment.  The spectra were taken
with Notre Dame's KN accelerator (E$_p$ = 1300 keV, top), Notre
Dame's JN accelerator (E$_p$ = 700 keV, middle), and the LUNA II
accelerator (E$_p$ = 130 keV, bottom).  The
$^{15}$N(p,$\gamma$)$^{16}$O ground state transition $\gamma$-ray
peak is indicated in each spectrum with an
arrow.\label{fig:sample_specta}}
\end{figure}

At the NSL, the $\gamma$-rays were observed using a HPGe Clover
detector, which consists of four HPGe crystals contained in the
same cryostat.  This unique arrangement allows the separate
detectors to be used in so called add-back mode~\cite{Dababneh04}.
At LUNA, a single crystal, 115\% HPGe detector from Bochum,
Germany was used for the detection of $\gamma$-rays.  Several
sample spectra are given in Figure~\ref{fig:sample_specta}.

The primary advantage of the LUNA II facility is the low
background environment created by the rock cover from the Gran
Sasso mountains.  The rock shields from cosmic rays and therefore
decreases the $\gamma$-induced background from cosmic rays in the
detector and has been shown~\cite{Bemmerer05} to suppress the
E$_\gamma$ \textgreater ~3.5 MeV background count rate in a Ge
detector by three orders of magnitude.  In each experiment the
detectors were set up at an angle of 45$^{\circ}$ with respect to
the beam direction, allowing the position of the detector to be
set as close as possible to the reaction position. The relative
efficiency of the $\gamma$-ray detector systems was measured using
radioactive sources along with well known capture $\gamma$
reactions.  At Notre Dame the relative efficiencies for high
energy $\gamma$-rays were established using the
$^{27}$Al(p,$\gamma$)$^{28}$Si resonances at 992
keV~\cite{Keinonen76} and 1183 keV~\cite{Meyer75} and the
$^{23}$Na(p,$\gamma$)$^{24}$Mg resonance at 1318
keV~\cite{Zijderhand90}.  The efficiency was extended to a
$\gamma$-energy of 12.79 MeV using the
$^{11}$B(p,$\gamma$)$^{12}$C reaction at 675 keV and 1388 keV
following the method of reference ~\cite{Zijderhand90}. At LUNA
the higher energy efficiencies were determined using the 278 keV
resonance in $^{14}$N(p,$\gamma$)$^{15}$O ~\cite{Imbriani05}] and
the 163 keV resonance in $^{11}$B(p,$\gamma$)$^{12}$C
~\cite{Ajzenberg86}. The $^{11}$B resonance has a small angular
distribution~\cite{Cecil1985}, and the correction for the relative
intensity is less than 3\%.

\begin{figure}[tbp]
\caption{Efficiency curves obtained at LUNA. Circles are
measurements at a distance of 1 cm, squares at 3 cm, and diamonds
at 20 cm.  Closed symbols represent data without summing
corrections, open symbols include summing corrections.
\label{fig:efficiancy_LUNA}}
\includegraphics[width=\figwidth]{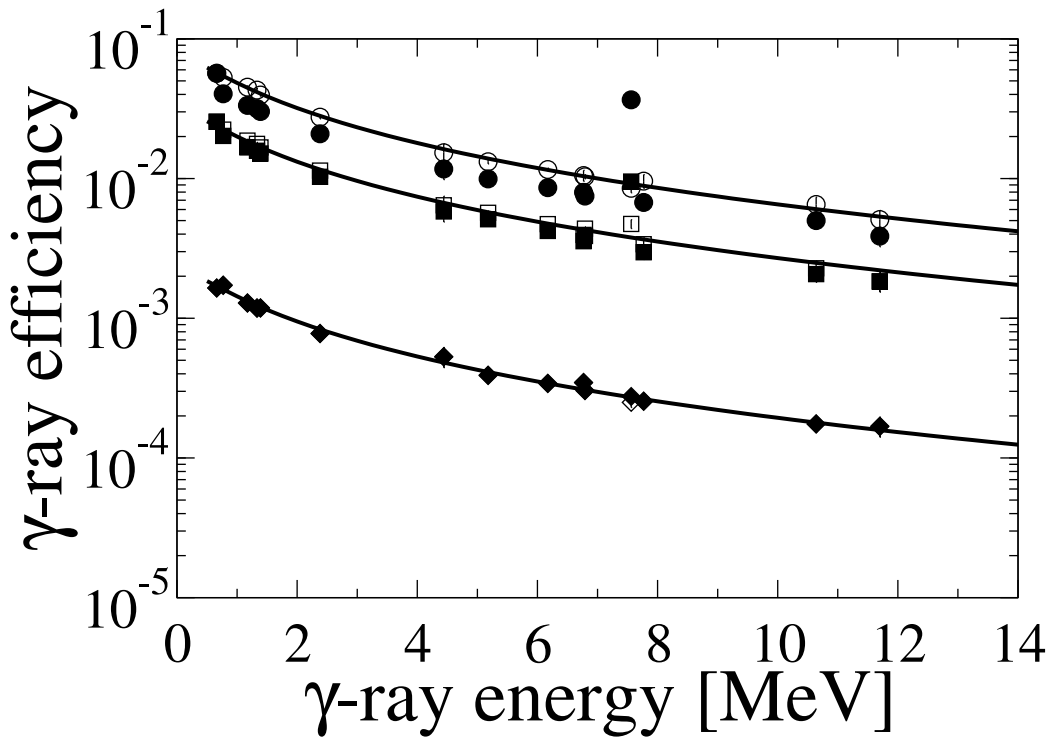}
\end{figure}

While summing is not of concern for the ground state transition of the
$^{15}$N(p,$\gamma$)$^{16}$O reaction itself, summing plays a
significant role in the determination of the $\gamma$-efficiency.
For this reason the efficiency measurements at both laboratories
were carried out at several different detector-target distances
and the data were simultaneously fitted for all distances
following the procedure described by Imbriani et
al.~\cite{Imbriani05} (see Figure~\ref{fig:efficiancy_LUNA}).  For
 the 1 cm and 5 cm measurements, there is one point which has
significant summing corrections.  This point corresponds to the ground
state transition in the E$_p$ = 278 keV $^{14}$N(p,$\gamma$)$^{15}$O
reaction.  While the ground state transition of this reaction has
a small branching ratio, each of the other cascades have significant
probabilities, which lead to strong summing effects~\cite{Imbriani05}.

\section{\label{sec:exp_results}\bfseries{Experimental Results}}

\subsection{\bfseries{Cross Section Determination}}

\begin{figure}[tbp]
\caption{Cross section for the ground state transition of the
$^{15}$N(p,$\gamma_0$)$^{16}$O reaction versus center of mass
energy.\label{fig:new_cs}}
\includegraphics[width=\figwidth]{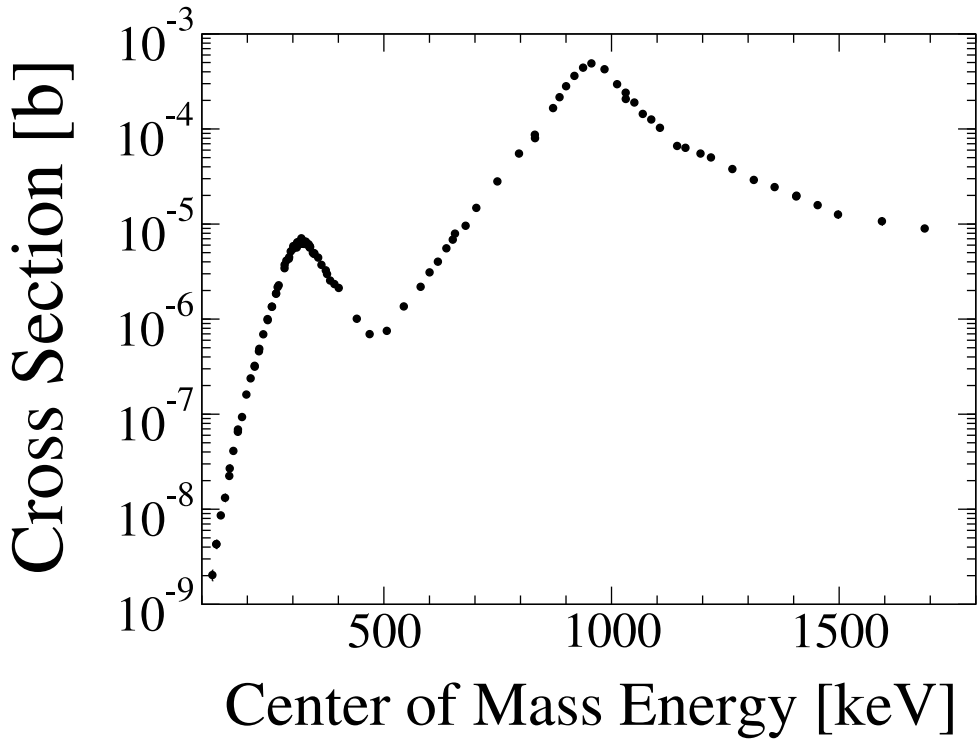}
\end{figure}

The excitation function for the ground state transition of the
reaction $^{15}$N(p,$\gamma$)$^{16}$O has been measured in the
energy range of 131 to 1800 keV. It consists of three distinct,
overlapping sections, LUNA data from 131 keV to 400 keV, JN data
from 285 keV to 700 keV, and KN data from 700 to 1800 keV. The
experimental yield Y (number of reactions per projectile) at proton
energy E$_p$ corresponds to the cross section $\sigma$(E) integrated
over the target thickness $\Delta$:

\begin{equation}
Y(E_p)= \int_{E_p-\Delta}^{E_p}\frac{\sigma(E)}{\epsilon(E)}dE =
\sigma(E_p)\int_{E_p-\Delta}^{E_p}\frac{f(E)}{\epsilon(E)}dE,
\label{eqn6}
\end{equation}

where $\epsilon$ is the stopping power~\cite{website:SRIM},
and f(E) represents the energy dependence of the cross section
in the integration interval:
\begin{equation}
\sigma(E) = \sigma(E_p)f(E),
\end{equation}

with $\sigma$(E$_p$) the cross section at E$_p$. This reduces to the well known
thin target yield equation if the cross section is constant over the target
thickness, f(E)=1. However, at energies where the cross section varies
significantly over the target thickness,the yield has to be
corrected to extract the cross section. This correction
factor is given by:

\begin{equation}
1/\int_{E_p-\Delta}^{E_p}\frac{f(E)}{\epsilon(E)}dE
\end{equation}
and requires the knowledge of the energy dependence of the cross
section in the energy interval $\Delta$, f(E). This problem can be
solved by an iterative method. In a first step f(E)=1 is assumed
resulting in an approximation of the cross section which in turn
can be used to calculate a new correction factor. This process is
continued until no change in the resulting cross section is
obtained. This process quickly converges after 2 to 3 iterations.
The resulting cross section is shown in Figure~\ref{fig:new_cs}.

An alternative method to determine the cross section was
used in which the data were analyzed following the approach described in
\cite{Imbriani05}.  Here, a brief review of the method is given, which
consists of the analysis of the shape of the $\gamma$-line.

The cross section, the stopping power of protons in TiN, the
$\gamma$-ray efficiency, and the detector resolution all have an
influence on the $\gamma$-ray line shape observed in a spectrum.
The observed energy of the $\gamma$-ray is related to the proton
energy by~\cite{Cauldrons88}

\begin{equation}
E_{\gamma} = E_p\frac{M}{M+m} + Q - E_x -\Delta E_{Rec} + \Delta E_{Dop}
\end{equation}

The number of counts N$_i$ in channel i of the $\gamma$-spectrum,
corresponding to the energy bin E$_{\gamma i}$ to E$_{\gamma i}$ +
$\delta$E$_\gamma$ ($\delta$E = dispersion in units of keV per
channel) is given by the expression

\begin{equation}\label{cts}
N_{i}~=~ \frac{\sigma(E_{pi})\delta
E_{\gamma}\eta_{fep}(E_{\gamma i})b_j }
{\epsilon(E_{pi})}
\end{equation}

for E$_{pi}\leq$ E$_p$ (E$_{pi}$ = proton energy corresponding to
channel i, E$_p$ = incident proton energy), where $\sigma(E_{pi})$
is the capture cross section, $\eta_{fep}(E_{\gamma i})$ is the
$\gamma$-ray detection efficiency, $\epsilon(E_{pi})$ is the
stopping power and b$_j$ is the branching of the associated decay.
The conversion from E$_{\gamma i}$ to E$_{pi}$ includes the
Doppler and recoil effects.  The result is folded with the known
detector resolution $\Delta$E$_\gamma$ to obtain the experimental
line-shape.

Finally, to infer the cross section in the energy window defined
by the target thickness, the cross section was written as the sum
of two resonance terms and a constant, non-resonant term as described
in~\cite{Rolfs73}.  In these fits, the free
parameters were the non-resonant astrophysical S factor and the
gamma-ray background parameters.  The results of both methods
are in excellent agreement.

The absolute cross section for the $^{15}$N(p,$\gamma$)$^{16}$O
reaction has been measured on top of the two broad resonances at
proton energies of 338 keV and 1028 keV using two independent well
known reaction standards and two independent methods.  All
measurements were performed at three distances (d = 1 cm, 5 cm, 20
cm) to check for systematic errors.  The exception to this
protocol was the $^{15}$N(p,$\gamma$)$^{16}$O measurement at E$_p$
= 338 keV, where the yield at 20 cm was too low.  In the first
method the cross sections were determined relative to the thick
target yield Y$_{\infty}$($^{27}$Al) of the well known
$^{27}$Al(p,$\gamma$)$^{28}$Si resonance at 992 keV:

\begin{equation}
Y_{\infty}(^{27}Al)=\frac{\lambda^2(E_R)}{2
\epsilon(E_R)}\frac{M+m}{M} \omega \gamma_R,
\label{eqn:thick_target_yield}
\end{equation}

with m (M) the mass of the projectile (target), $\lambda$ the
DeBroglie resonance wave length and $\omega \gamma_R$ the
resonance strength.  Adopting the resonance strength of 1.93 $\pm$
0.13 eV from~\cite{Paine&Sargood79,Antilla77,Keinonen76} the cross
sections values of $\sigma_{338}$=6.7 $\pm$ 0.6 $\mu$b and
$\sigma_{1028}$=446 $\pm$ 40 $\mu$b are obtained. The errors include the
statistical error ($\le$ 3 $\%$), the error on the relative
efficiency (2 $\%$), the error on the relative charge measurement
(2$\%$), the uncertainty on stopping power values (5$\%$), and the
uncertainty of 7 $\%$ arising from the reference resonance strength of
$^{27}$Al.

In the second method the cross sections were determined relative
to the well known 429 keV resonance in
$^{15}$N(p,$\alpha_1\gamma$)$^{12}$C~\cite{Ajzenberg86}. In this
case the cross section was calculated relative to the integral
 over the yield curve of the reference resonance, A$^{ref}$.  The resonance
 strength $\omega\gamma$ is related to A$^{ref}$ by:
\begin{equation}
\omega\gamma=\frac{2}{\lambda^2}\frac{1}{n_t}A^{ref},
\label{eqn:a_ref}
\end{equation}
with n$_t$ the number of target atoms per cm$^2$ which can be calculated
from the measured target thickness and the stopping power values.  Using this
relationship for A$^{ref}$ (Equation~\ref{eqn:a_ref}), along with Equation~\ref{eqn:thick_target_yield}
and the thin target yield approximation for the $^{15}$N(p,$\gamma$)$^{16}$O
reaction, the following formula can be derived:

\begin{eqnarray}
\sigma_{(p,\gamma)}(E_p)&=&\left(\frac{\lambda^2(E_R)}{2}\right)\left(\frac{M+m}{M}\right) \\ \nonumber
&&\times \left(\frac{\eta(4.4MeV)}{\eta(12MeV)}\right)\left(\frac{Y(E_p)}{A^{ref}}\right)
\omega \gamma_R, \label{eqn:sigma_wg}
\end{eqnarray}

with $\eta$ the relative $\gamma$-ray efficiency. In this approach
the result is independent of all target uncertainties. The yield of
the 4.4 MeV $\gamma$-rays from the $^{15}$N(p,$\alpha_1\gamma$)$^{12}$C
was corrected for the well known angular distribution
~\cite{Barnes52,Kraus53}.  Using the resonance strength of
$\omega\gamma_R$=21.2 $\pm$ 1.4 eV from~\cite{Becker95}, values of
$\sigma_{338}$=6.4 $\pm$ 0.6 $\mu$b and $\sigma_{1028}$=436 $\pm$ 44 $\mu$b
are obtained. The errors include the statistical error ($\le$ 3 $\%$), the
error of the relative efficiency (2 $\%$), the error of the relative charge
measurement (2$\%$), the error of A$_{ref}$ owing to the numerical integration
and angular distribution correction (6$\%$) and the uncertainty of
7 $\%$ of the reference value~\cite{Becker95}.

The weighted average of the results from the two methods gives
final values of $\sigma_{338}$ = (6.5 $\pm$ 0.3) $\mu$b (5\%) and
$\sigma_{1028}$ = (438 $\pm$ 16) $\mu$b (4\%) taking into account
common error sources. The present results are compared to previous
values~\cite{Hebbard60,Rolfs74,Brochard73} in Table~\ref{table:sigma_summary}
 and are in very good agreement. An exception is the cross section
at 338 keV from Rolfs and Rodney~\cite{Rolfs74} which is about
50\% higher than the other values.  This discrepancy will be
discussed in more detail in the next section.  Data from LUNA and
the Notre Dame JN accelerators were normalized to the low energy
resonance, while data from the Notre Dame KN accelerator were
normalized to the higher energy resonance. Good agreement between
the data sets was found at the overlapping energies around 700
keV.

\begin{table}[tbp]
\begin{tabular}{c|cccc} \hline \hline
E$_R$ & Present & Rolfs & Brochard & Hebbard \\
& & \cite{Rolfs74} & \cite{Brochard73} & \cite{Hebbard60} \\
unit:[keV] & [$\mu$b] & [$\mu$b] & [$\mu$b] & [$\mu$b] \\
\hline
338  & 6.5 $\pm$ 0.3 & 9.6 $\pm$ 1.3 & 6.3* & 6.5 $\pm$ .3$^\dagger$ \\
1028 & 438 $\pm$ 16 & 420 $\pm$ 60  & 490* &  - \\
\hline \hline
\end{tabular}
\caption{Summary of present resonance cross sections in comparison
to previous results.  *No uncertainty is given. $^\dagger$ error
derived from given reproducibility of
5\%.\label{table:sigma_summary}}
\end{table}

\subsection{\bfseries{Branching Ratios}}

In previous experiments small cascade transitions have been
observed for the broad resonances at 338 keV and 1028
keV~\cite{Ajzenberg86}. For the 338 keV resonance a branching ratio
of (1.2 $\pm$ 0.1) $\%$ for the transition to the 6050 keV level was
found, in good agreement with the literature value of 1.2 $\pm$ 0.4
$\%$~\cite{Ajzenberg86}. For the 1028 keV resonance, branches to the
6050 keV level (1.0 $\pm$ 0.3) $\%$ and to the
7118 keV level (2.8 $\pm$ 0.3) $\%$ were measured, and were also in good
agreement with literature value of (1.4 $\pm$ 0.4) $\%$ and (3.1
$\pm$ 0.8) $\%$, respectively~\cite{Ajzenberg86}.

\subsection{\bfseries{Angular Distributions}}
In previous analyses of the $^{15}$N(p,$\gamma_0$)$^{16}$O
reaction~\cite{Rolfs74,Barker08b,MBB08}, a direct capture
component was required to fit the data.  In each of these cases,
only the l$_i$ = 0 initial wave was included in the calculation.
This component represents s-wave capture, and therefore interferes
with the resonant component of the cross section.  The s-wave
components yield isotropic $\gamma$-ray
distributions~\cite{Rolfs73,Rolfs74}.  However, an $l_i$ = 2
direct capture component is allowed for E1 transitions, and can
contribute to the reaction, possibly introducing some anisotropy
in the gamma decay.  In looking for this signature, the set-up was
tested using the 1 MeV resonance, which is reported to be
isotropic~\cite{Rolfs74}. Measurements were made with the central
Clover axis at several angles relative to the beam direction
with a nominal distance of 5 cm.  To extract additional data
points, the add-back feature was used to treat the left and right
halves of the Clover system as separate detectors, where Monte
Carlo calculations using the code GEANT4~\cite{Geant03} yielded an
effective angular offset of each half from the central detector
axis of 15$^{\circ}$.  Measurements with a calibrated $^{137}$Cs
source, and the isotropic 278 keV resonance of
$^{14}$N(p,$\gamma$)$^{15}$O made corrections to the absorption
effects possible. The results confirmed the isotropy of the
angular distribution of the ground state decay $\gamma$-rays from
the 1028 keV resonance in $^{15}$N(p,$\gamma_0$)$^{16}$O.

In the search for the l$_f$ = 2 component, angular distribution
measurements of $^{15}$N(p,$\gamma_0$)$^{16}$O were made at a
proton energy of 540 keV at detection angles of 0$^{\circ}$,
45$^{\circ}$, and 90$^{\circ}$ relative to the beam axis.  With
two Clover segments, this resulted in 6 data points. Two of these
points were excluded due to large absorption effects from the
shape of the target chamber.  The resulting angular distribution
was fit with a function of the form W($\theta$) =
a$_0^{exp}$ + a$_2^{exp}$ P$_2$(cos$\theta$), where P$_2$ is the L
= 2 Legendre polynomial, and a$_2^{exp}$= Q$_2$ a$_2$, where Q$_2$
is the usual geometrical correction factor due to the finite size of the
detector~\cite{Ferguson65}. The Q$_2$ term can be directly
determined using the $^{16}$O(p,$\gamma$) DC $\rightarrow$ 495 keV
reaction~\cite{Rolfs73}. This reaction has a well known angular
distribution of the form $W(\theta) = sin^2(\theta)$. Plotting the
measured W$(\theta)$ with respect to $sin^2(\theta)$, the
deviation of the slope from 1.0 gives directly the $Q_2$ value.
Using the left and right Clover halves to measure the
$^{16}$O(p,$\gamma$)$^{17}$F DC$\rightarrow$495 keV a value of
Q$_2$ = 0.975$\pm$0.020 was found. This results in an a$_2$ value
for the $^{15}$N(p,$\gamma_0$)$^{16}$O reaction at 540 keV of 0.08 $\pm$
0.10, which is consistent with isotropy.

\section{\label{sec:rmatrix}\bfseries{R-Matrix Analysis}}

\begin{figure}[ht]
\includegraphics[width=\figwidth]{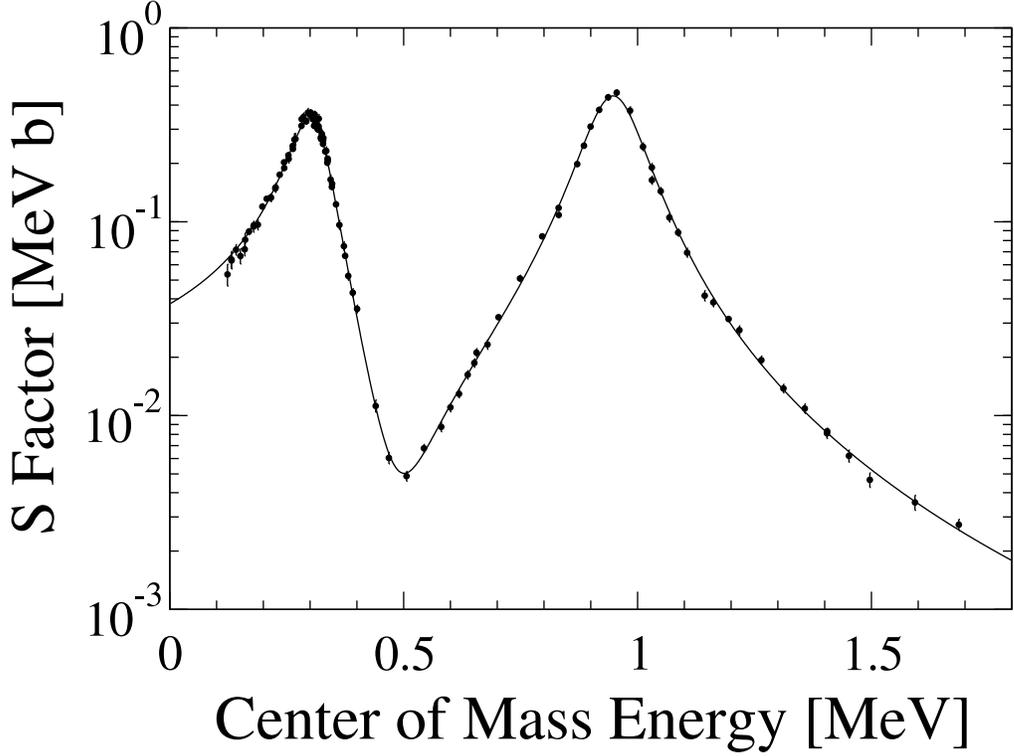}
\caption{\label{fig:my_data_wcalc} Experimental S factor for the
ground state transition of the $^{15}$N(p,$\gamma_0$)$^{16}$O
reaction shown together with the best fit results of an R-matrix
calculation using the code AZURE.}
\end{figure}

An R-matrix analysis was performed which mirrored the procedure of
previous analyses~\cite{Rolfs74,Barker08b,MBB08}. This analysis
included the two broad 1$^{-}$ resonance levels and a direct
capture contribution.  The analysis was performed with the
multi-channel R-matrix code AZURE~\cite{Azuma10}. Details of the
theory and the nomenclature are given in~\cite{Azuma10} and
references therein. The best fit results can be seen in
Figure~\ref{fig:my_data_wcalc} where the cross sections have been
converted to the astrophysical S-factor. This fit results in an
S(0) value of 39.6 keV b.

\begin{figure}[tbp]
\includegraphics[width=\figwidth]{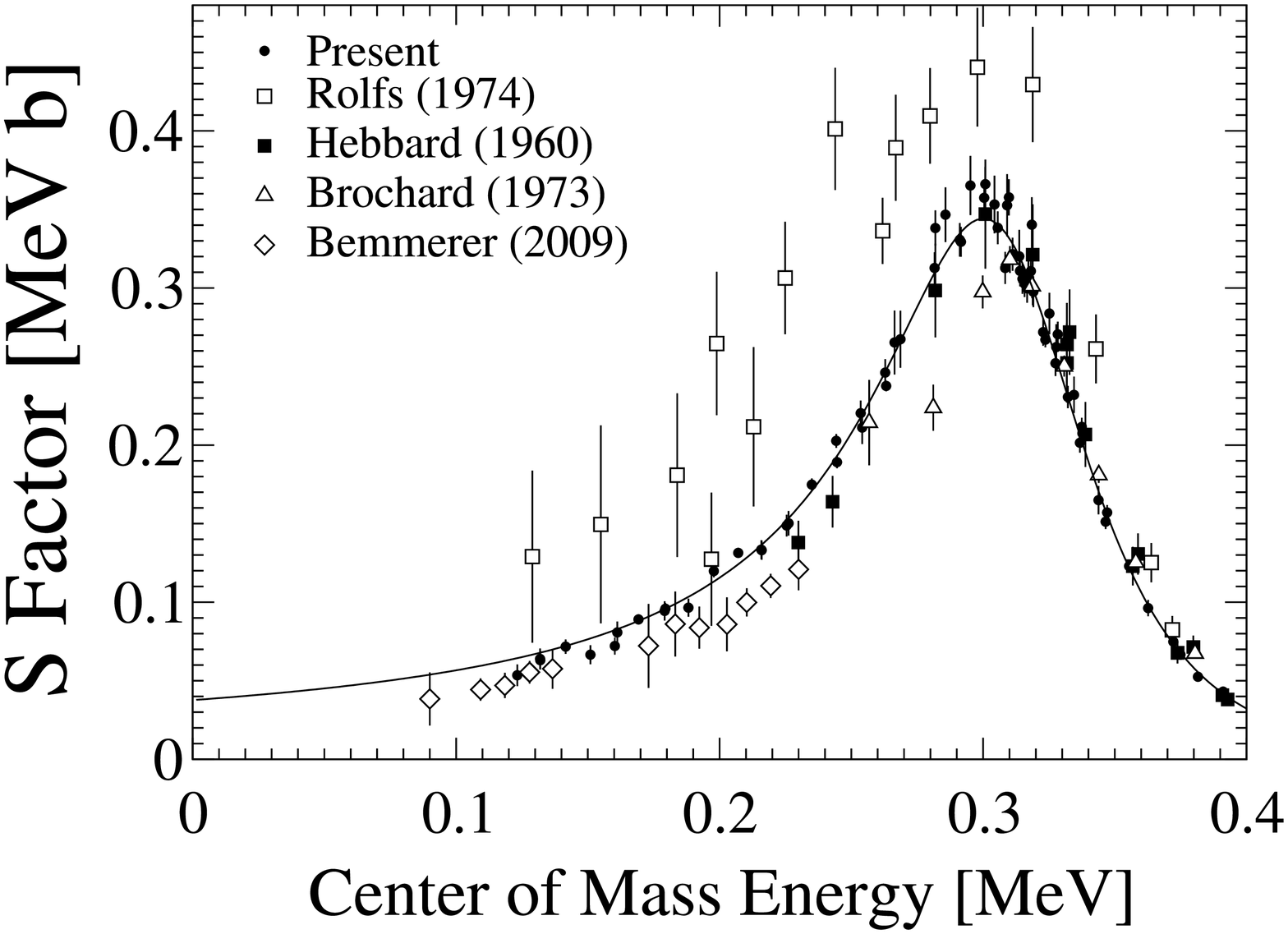}
\caption{\label{fig:all_data_wcalc_zoom} Low energy region of
present data set (filled circles) shown with AZURE R-matrix
calculation along with previous measurements of
Rolfs~\cite{Rolfs74}(open squares),
Hebbard~\cite{Hebbard60}(filled squares),
Brochard~\cite{Brochard73}(open triangles), and
Bemmerer~\cite{Bemmerer09}(open diamonds) in the region of 0 - 400
keV}
\end{figure}

Looking closer at the low energy region (see
Figure~\ref{fig:all_data_wcalc_zoom}) and comparing the results to
previous experiments, the low energy data of~\cite{Rolfs74} are
inconsistent with the present results.  While this data set agrees
very well at higher energies it starts to deviate below a proton
energy of 400 keV.  There is no obvious reason for this
inconsistency but it should be noted that the low energy data of
Rolfs and Rodney~\cite{Rolfs74} carry a significant uncertainty in
this region. The data of Hebbard~\cite{Hebbard60} are in
reasonably good agreement above 230 keV. The data of
Brochard~\cite{Brochard73} show good agreement except for the
three data points below the 338 keV resonance which show a large
scatter.The recent Bemmerer data~\cite{Bemmerer09} are
systematically lower than the present data, and also seem to have
a slightly different energy dependence.

The parameters for the best fit are given in
Table~\ref{table:15N_param2}. With the exception of the first
resonance, most parameters are in fair agreement with previous
results~\cite{MBB08,Barker08b}. The width of the first resonance
is dominated by the alpha width~\cite{Ajzenberg86}. This parameter is,
therefore, well constrained by the present (p,$\gamma$) data.  The
resonance strength, however, is determined by the product of the proton
and $\gamma$ width. Without including proton scattering data in the
fit, these parameters are not well constrained. However, it should
be noted that tests showed that this ambiguity has no influence on
the extrapolation of the data. In these tests, the proton width
($\gamma_{p_1}$) was fixed at different values while allowing the
other parameters to vary.  For the higher energy resonance the
proton and alpha width are comparable, which provides more of a
constraint on the parameters. In addition, the present result for
the ground state ANC is significantly larger than the value of
\cite{MBB08} which cannot be attributed to the choice of radius
(see below). While these differences do not have an impact on the
extrapolation of the S-factor to lower energies, a more thorough
analysis is warranted for the interpretation of the R-matrix
parameters. This will be addressed in a forthcoming
publication~\cite{LeBlanc10} where the results of simultaneous
multi-channel fits to all relevant reaction channels will be
presented.

\begin{table}[tbp]
\begin{tabular}{c|cccccccccccc} \hline \hline
 Reference & E$_1$ & $\gamma^2_{p_1}$ & $\gamma^2_{{\alpha_0}_1}$ &
$\gamma^2_{\gamma_1}$(int) & $\Gamma_{\gamma_1}$ & E$_2$ &
$\gamma^2_{p_2}$ & $\gamma^2_{{\alpha_0}_2}$ &
$\gamma^2_{\gamma_2}$(int) &$\Gamma_{\gamma_2}$ &
C$_{g.s}$[fm$^{1/2}$] & $\theta_{g.s}$  \\
    & J$^{\pi}$ = l$^-$ & l$_p$ = 0 & l$_{\alpha}$ = 1 &  &  &
J$^{\pi}$ = l$^-$ & l$_p$ = 0 & l$_{\alpha}$ = 1 &  &  & s,d$\rightarrow$p & \\
\hline
Present     & 12.438 & 52.8  & 13.5 & 51.3  & 33.8    & 13.087 & 309.1 & 5.0   & 34.1 & 38.7  & 23.22  & 0.608 \\
\cite{MBB08} & 12.439 & 280.9 & 12.5 & - & 8.8$\pm$1.5 & 13.089 & 271.4 & 6.1   &  -  & 50$\pm$8 & 13.85  & -     \\
\cite[RR]{Barker08b} & 12.452 & 93.6  & 13.5 & 38.0  &  -    & 13.111 & 416.0 & 0.812 & 0.5  & -    & -      & 1.944 \\
\cite[HH]{Barker08b} & 12.447 & 355.2 & 10.6 & 7.2  &  -      & 13.087 & 265.2 & 5.4   & 56.2 & -     & -      & 0.569 \\
\hline
\end{tabular}
\caption{\label{table:15N_param2} Best fit R-matrix parameters for
the $^{15}$N(p,$\gamma$)$^{16}$O reaction (E$_x$ in MeV, $\gamma^2_i$ in
keV). Channel radii of $a_p=5.030$ fm and $a_{\alpha} = 6.500$ fm
were used in the fit. The boundary condition BC in the fit was
chosen to be equal to the shift function at the energy of the
lowest J$^{\pi}$=1$^-$ resonance. Because of this choice of BC,
the $\Gamma_{\gamma_1}$ (in eV) can be calculated directly from
the reduced width amplitude while $\Gamma_{\gamma_2}$ is
calculated from the Barker transformation of the reduced width
amplitude~\cite{Azuma10}. The parameters shown in this table are
mainly intended to enable the reader to reproduce the fit
presented in this manuscript, and should not be regarded as final.
With respect to the physical interpretation of these fit
parameters, a detailed discussion will be presented in a
forthcoming paper~\cite{LeBlanc10}.}
\end{table}

\begin{figure}[hbt]
\begin{center}
\subfigure[]{\label{fig:chi2_vs_rp}\includegraphics[width=0.45\columnwidth]{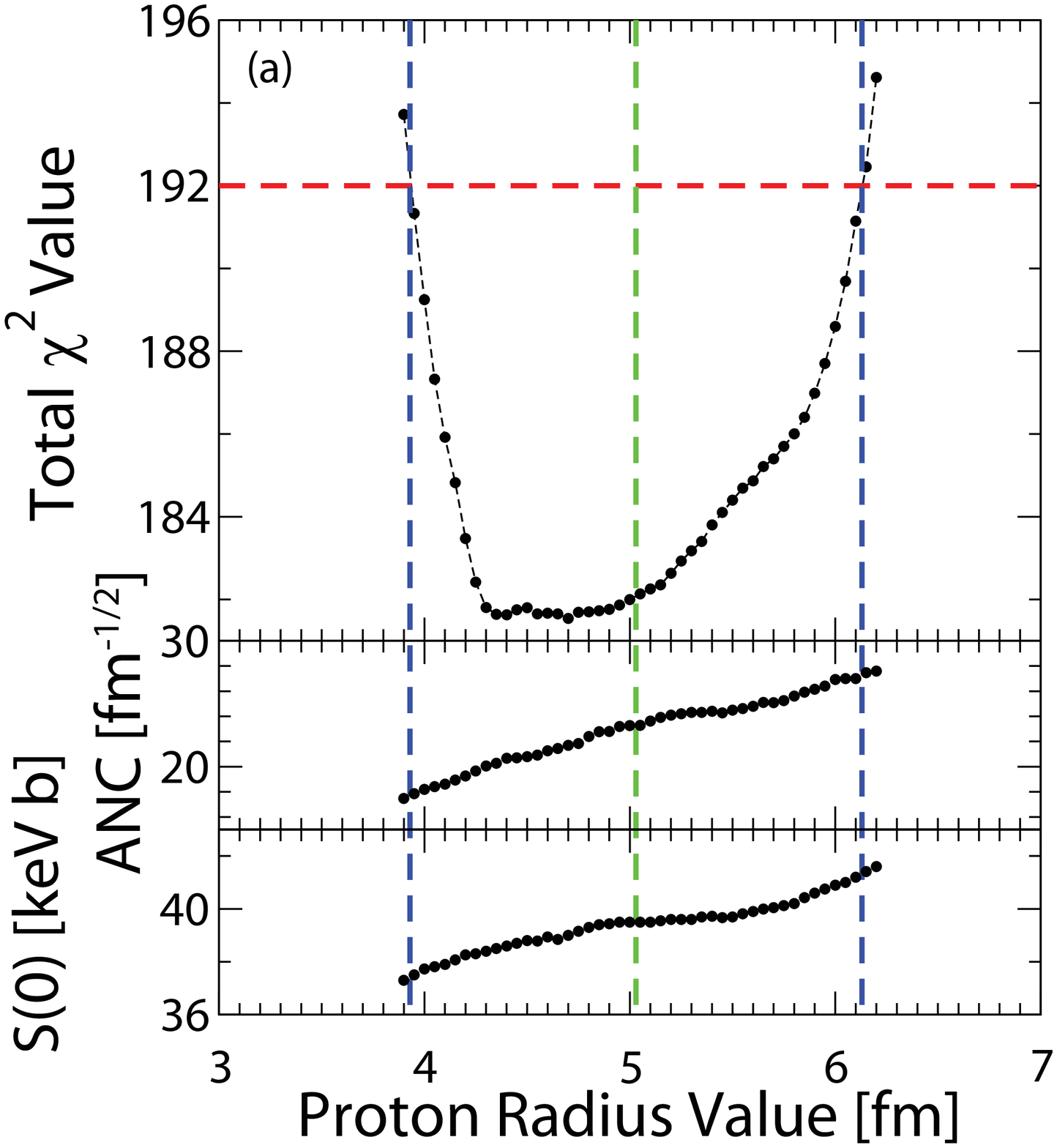}}
\subfigure[]{\label{fig:chi2_vs_ANC}\includegraphics[width=0.45\columnwidth]{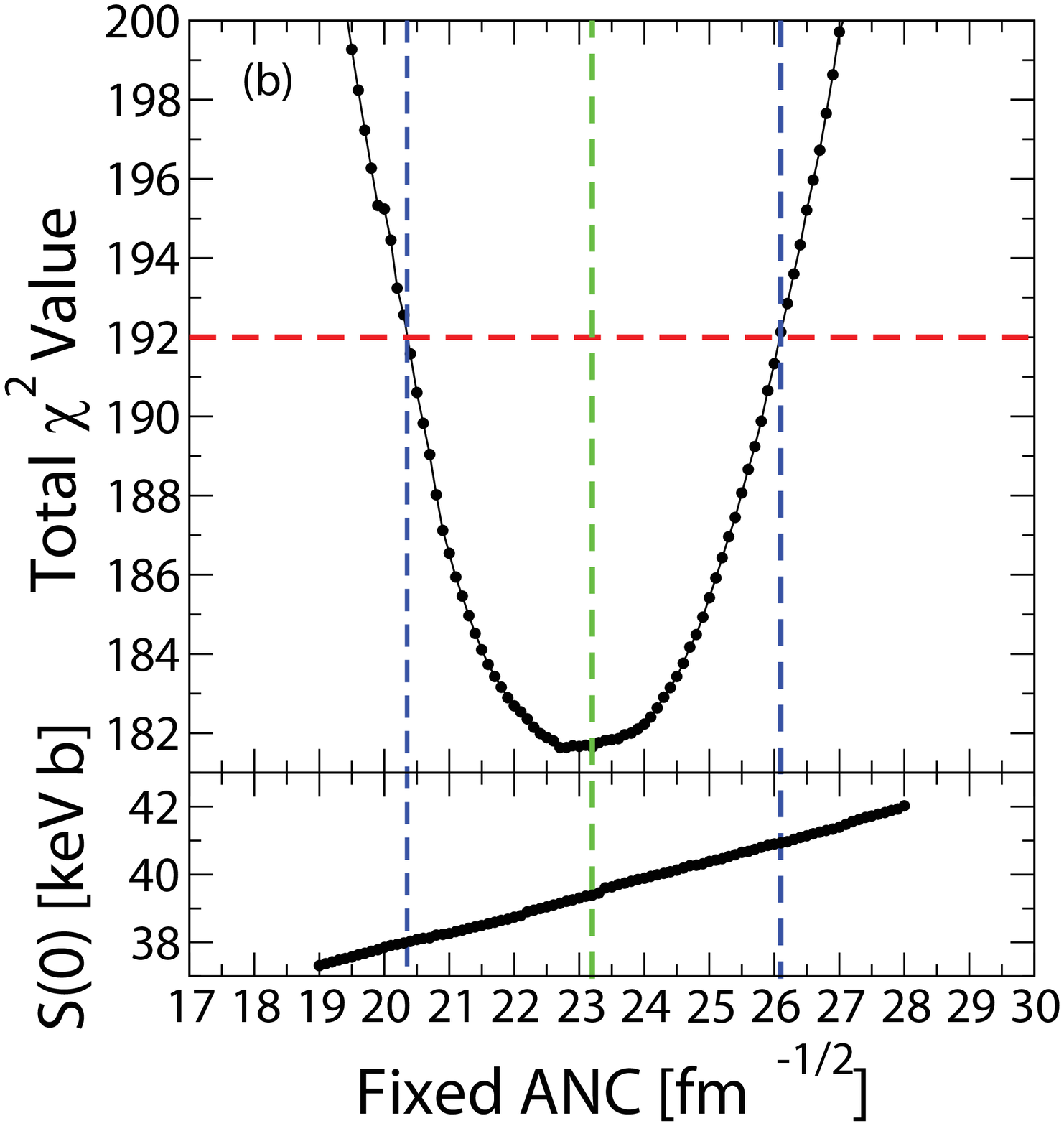}}
\end{center}
\caption{Dependence of AZURE fits on key parameters used in the
fits. Parameters were varied noting that for a 9 parameter fit, a
70\% confidence region is defined for $\approx$ $\chi^2$ =
$\chi^2_{min}$ + 10~\cite{Jam94}.  Even though a lower $\chi^2$
was found with a radius of about 4.4 fm, 5.03 was chosen so as
to be more comparable to~\citet{Barker08b}.  ANC was found to be
(23 $\pm$ 3) fm$^{-1/2}$.  Fixing the ANC to the value found
in~\cite{MBB08} gives a $\chi^2$ that is much higher than the
current best fit value.\label{fig:param_dep}}
\end{figure}

In the present parameter space the only non-s-wave contribution arises from the d-wave component of the direct capture. Using the parameters for the best fit the $\gamma$-ray angular distribution was calculated at the minimum of the cross section between the two resonance at 540 keV yielding a value of a$_2$ = 0.034. The result is consistent with the experimental upper limit of 0.18 (see Section~\ref{sec:exp_results} C).

The sensitivity of the best fit was tested against several key
parameters.  In multi-parameter fitting, uncertainties of
specific parameters are determined in terms of confidence regions.
For a nine parameter fit (E$_{\lambda}$,$\gamma_i$ for both levels
plus ANC or $\theta_{g.s.}$) a 70\% confidence region for one of the
parameters is defined by the range where $\chi^2$ $\leq$ $\chi^2_{min}$
+ 10~\cite{Jam94}.  The degrees of freedom is 102 (113 data points, 9
parameters) resulting in a reduced $\chi^2$ of 1.8.

Radii of r$_p$ = 5.03 fm, and r$_{\alpha}$ = 6.5 fm, were taken
from~\cite{Barker08b}. The dependence of the fit on the proton
radius was tested, and the results are given in
Figure~\ref{fig:chi2_vs_rp}. Any proton radius value between 4 and
6 fm is considered acceptable.  Variation of the radius over this
range corresponds to only a 5\% uncertainty in the S(0)
extrapolation. This test also showed that the ANC is not very
sensitive to the choice of the radius. The best fit gives a ground
state ANC of (23 $\pm$ 3) fm$^{-1/2}$ (corresponding to a reduced
width amplitude of $\theta_{g.s.}$ = 0.61).  The error
associated with the ANC was determined by fixing the ANC at
different values, and finding a new best fit. The results of this
procedure can be seen in Figure~\ref{fig:chi2_vs_ANC}. Even though
the results give a 13\% uncertainty for the ANC, the variation in
S(0) from this procedure is only 4\%.

\begin{figure}[htb]
\includegraphics[width=\figwidth]{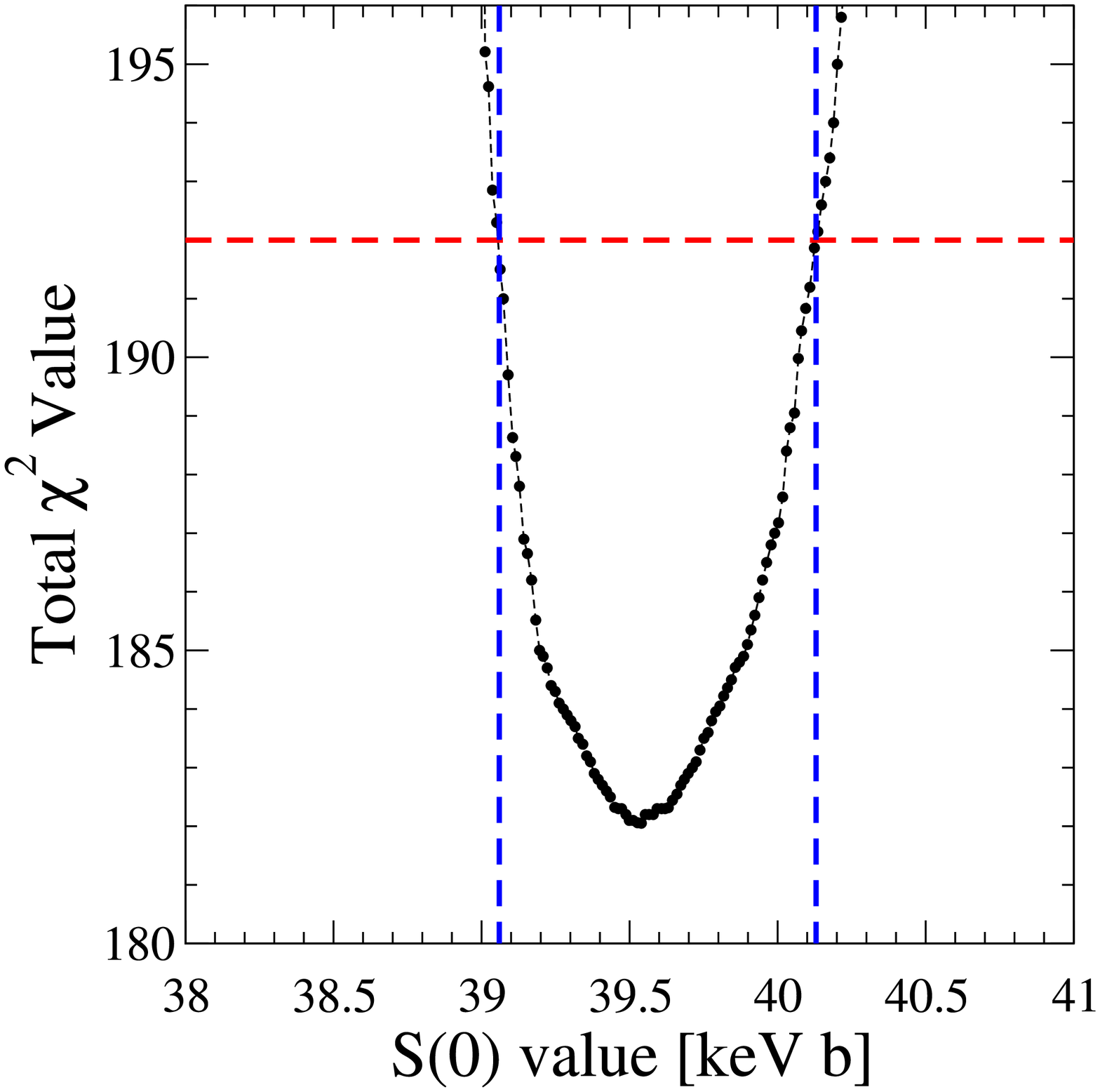}
\caption{\label{fig:chi2_vs_S0} Variation of $\chi^2$ with extrapolated
S(0) values (for details, see text).}
\end{figure}

To evaluate the uncertainty of the fit itself, fits were performed
which were forced to result in different S(0) values by inserting
a fake data point at 1.5 keV with extremely small errors.  The
small error of this data point forces the calculation to match
this fictitious data point and thus vary the extrapolation. By
varying the fictitious point and observing the change in the
$\chi^2_{tot}$, the experimental uncertainty of the extrapolation
can be determined (see Figure~\ref{fig:chi2_vs_S0}). This results
in an uncertainty of $\pm$ 0.6 keV b. Including the 5\% error of
the absolute cross section gives a final S(0) value of (39.6 $\pm$
2.6) keV b. This value is compared in Table
\ref{table:sfactor_summary} with previous extrapolations
\cite{Hebbard60,Rolfs74,Simpson06,Barker08b,MBB08}. The present
result is in good agreement with the original value of Hebbard
\cite{Hebbard60}, the Barker \cite{Barker08b} analysis of the
Hebbard data (labelled HH in Table \ref{table:sfactor_summary},
and the results of Mukhamedzhanov et al. \cite{MBB08}. The
extrapolation of the Rolfs and Rodney data \cite{Rolfs74} (Rolfs
and Rodney \cite{Rolfs74}, Simpson \cite{Simpson06} and Barker
\cite{Barker08b}, labelled RR in Table
\ref{table:sfactor_summary}) are all significantly higher because
of the larger low energy cross sections used for their fit
analysis.

\begin{table}[htp]
\caption{Summary of Previous Results for extrapolated S(0) values\label{table:sfactor_summary}}
\centering
\begin{tabular}{l|c}\hline \hline
Analysis      & $S(0)_{\gamma}$ (keV b) \\ \hline
Hebbard 1960 \cite{Hebbard60}       &     32      \\
Rolfs 1974   \cite{Rolfs74}      & 64 $\pm$ 6  \\
Barker 2008 (RR) \cite{Barker08b}  & $\approx$ 50-55 \\
Barker 2008 (HH) \cite{Barker08b}  & $\approx$ 35 \\
Mukhamedzhanov   \cite{MBB08} & 36.0 $\pm$ 6.0   \\
Present            & 39.6 $\pm$ 2.6 \\
\hline \hline
\end{tabular}
\end{table}

\section{\bfseries{Reaction Rate}}

Using the results from the AZURE extrapolations, the reaction
rates can be numerically determined using the formalism outlined
in~\cite{Angulo99}

\begin{eqnarray}
N_A\langle\sigma v\rangle &=& 3.73 \cdot 10^7 \mu^{-\frac{1}{2}}
T^{-\frac{3}{2}} \\ \nonumber
 && \times \int_0^{\infty}S(E)e^{(-2\pi\eta)}e^{(-11.605E/T)}dE
\end{eqnarray}

with $\mu$ the reduced mass, T the temperature in GK, and $\eta$
the Sommerfeld parameter; the results are in cm$^3$ mole$^{-1}$
s$^{-1}$. The above function was numerically integrated for
temperatures between T = 0.01 - 10 GK. The present data covers an
energy range of E = 0 to 1.8 MeV, which validates the integration
up to a temperature of T = 1 GK. For higher temperatures, the
S-factor curve must be extended to higher energies beyond what the
present experimental data cover. We followed the procedure of the
NACRE compilation~\cite{Angulo99} which handles this difficulty by
equating all higher energy S-factor values with the highest energy
data point where the cross section varies only slowly with energy.
The results are given in Table~\ref{table:react_rate_values}. The
present rate is at lower temperatures up to a factor two lower
than previous rates reflecting the change in the low energy
S-factor.

\begingroup
\squeezetable
\begin{table}[tbh]
\begin{tabular}{cccc} \hline \hline
T9 [GK] & N$_A\langle \sigma v \rangle_{Present}$ & N$_A\langle
\sigma v \rangle_{NACRE}$ & N$_A\langle \sigma v \rangle_{CA88}$ \\
& & \cite{Angulo99} & \cite{CaF88}\\
 \hline
0.010   &   2.284E-21   &   4.33E-21    &   3.93E-21\\
0.015   &   1.377E-17   &   2.66E-17    &   2.34E-17\\
0.020   &   3.322E-15   &   6.50E-15    &   5.61E-15\\
0.030   &   3.215E-12   &   6.41E-12    &   5.38E-12\\
0.040   &   2.456E-10   &   4.96E-10    &   4.08E-10\\
0.050   &   5.378E-09   &   1.09E-08    &   8.90E-09\\
0.060   &   5.681E-08   &   1.16E-07    &   9.36E-08\\
0.070   &   3.747E-07   &   7.64E-07    &   6.14E-07\\
0.080   &   1.784E-06   &   3.63E-06    &   2.91E-06\\
0.090   &   6.699E-06   &   1.36E-05    &   1.08E-05\\
0.100   &   2.104E-05   &   4.23E-05    &   3.37E-05\\
0.150   &   1.280E-03   &   2.46E-03    &   1.92E-03\\
0.200   &   1.908E-02   &   3.52E-02    &   3.09E-02\\
0.300   &   6.143E-01   &   1.05E+00    &   1.57E+00\\
0.400   &   4.682E+00   &   7.48E+00    &   1.45E+01\\
0.500   &   1.668E+01   &   2.58E+01    &   5.42E+01\\
0.600   &   3.908E+01   &   5.90E+01    &   1.27E+02\\
0.700   &   7.213E+01   &   1.07E+02    &   2.25E+02\\
0.800   &   1.165E+02   &   1.68E+02    &   3.39E+02\\
0.900   &   1.763E+02   &   2.46E+02    &   4.67E+02\\
1.000   &   2.599E+02   &   3.50E+02    &   6.18E+02\\
1.500   &   1.485E+03   &   1.67E+03    &   2.33E+03\\
2.000   &   4.735E+03   &   5.05E+03    &   6.61E+03\\
3.000   &   1.489E+04   &   1.55E+04    &   1.97E+04\\
4.000   &   2.428E+04   &   2.27E+04    &   3.12E+04\\
5.000   &   3.064E+04   &   2.61E+04    &   3.83E+04\\
6.000   &   3.435E+04   &   2.86E+04    &   4.19E+04\\
7.000   &   3.624E+04   &   3.06E+04    &   4.31E+04\\
8.000   &   3.699E+04   &   3.23E+04    &   4.29E+04\\
9.000   &   3.706E+04   &   3.37E+04    &   4.19E+04\\
10.000& 3.674E+04   &   3.51E+04    &   4.04E+04\\
\hline
\end{tabular}
\caption{\label{table:react_rate_values} Table of reaction rates
for $^{15}$N(p,$\gamma_0$)$^{16}$O.  Present rates, along with the
NACRE~\cite{Angulo99}, and CA88~\cite{CaF88} results are given.}
\end{table}
\endgroup

Following the example of the NACRE compilations, the above results
were fit using the following parametrization
\begingroup
\begin{eqnarray}
N_A <\sigma v> &=& a_110^9T^{-\frac{3}{2}}exp[a_2 T^{-\frac{1}{3}}-(T/a_3)^2] \\ \nonumber
 & & [1 + a_4T + a_5T^2] + a_610^3 T^{-\frac{3}{2}}\\\nonumber
 & & exp(a_7/T) + a_810^6T^{-\frac{3}{2}}exp(a_9/T),
\end{eqnarray}
\endgroup
where the best fit parameters can be found in Table~\ref{table:rr_fit_par}.

\begin{table}[tbh]
\begin{tabular}{ccc}
\hline \hline
a$_1$ =   0.523  & a$_4$ = 6.339  & a$_7$ = -2.913  \\
a$_2$ = -15.240  & a$_5$ = -2.164 & a$_8$ =  3.048 \\
a$_3$ =   0.866  & a$_6$ = 0.738  & a$_9$ = -9.884 \\
\hline \hline
\end{tabular}
\caption{\label{table:rr_fit_par} Best fit parameters for the
$^{15}$N(p,$\gamma_0$)$^{16}$O reaction rate using the NACRE fitting formulations.}
\end{table}


\section{\bfseries{Conclusion}}
The result of this work clearly demonstrates that there are
significant uncertainties in the low energy cross section data of
radiative capture processes of astrophysical relevance, despite
many decades of low energy reaction studies. These uncertainties
affect directly our understanding and interpretation of solar and
stellar hydrogen burning phenomena. In this case the new results
influence primarily the leakage rate from the CN to the ON cycle
in stellar burning via the $^{15}$N(p,$\gamma$)$^{16}$O radiative
capture process, which is reduced by a factor of two compared to
the previous rate used traditionally in CNO nucleosynthesis
simulations. In particular, the change in rate will modify the
equilibrium abundance of $^{16}$O, which is correlated with the
leakage rate of $^{15}$N(p,$\gamma$)$^{16}$O from the CN cycle and
the rate of $^{16}$O(p,$\gamma$)$^{17}$F in the NO cycle. However
a detailed study of the astrophysical impact of the present
measurement goes beyond the aim of the present work, but should
benefit from recent studies of low energy reaction
rates~\cite{Azuma10}.

The reliability of stellar reaction rates depends critically on
the quality of the experimental cross section data. Direct
measurements of the reaction cross sections at the Gamow range of
stellar burning have been successful in only a few cases of
reactions between light nuclei. Considering the anticipated count
rates for the CNO radiative capture reactions we will continue to
rely for most cases on the extrapolation of low energy
measurements into the Gamow range. The present analysis clearly
demonstrates that this requires a two-fold approach, pursuing the
direct reaction measurements to the lowest possible energies in a
background shielded environment but also expanding the
experimental range of the measurements to determine unambiguously
the various reaction components of the radiative capture process.
The latter step is essential for minimizing the uncertainties in
the R-matrix analysis of the cross section and can be complemented
by independent studies which explore independently the strength of
specific "hidden" reaction components such as the direct capture
through ANC measurements and analysis.

The approach taken here for the study of the
$^{15}$N(p,$\gamma$)$^{16}$O reaction has succeeded in combining
both the efforts of improving on the extent and quality of the low
energy cross section data in underground accelerator experiments.
At the same time the study has improved on the detailed
measurement of higher energy data providing a better constraint on
determining the external capture component and its impact on the
low energy extrapolation of the reaction cross section. The
combination of these two complementary measurements successfully
reduced the overall uncertainty in the
$^{15}$N(p,$\gamma$)$^{16}$O reaction rate.

We are extremely grateful for the help of the technical staff of
both the Nuclear Science Laboratory at the University of Notre
Dame and that of the Gran Sasso facility.  REA thanks the NSERC
for partial financial support through the DRAGON grant at TRIUMF.
This work was funded in part by the National Science Foundation
through grant number 0758100, the Joint Institute for Nuclear
Astrophysics grant number 0822648, along with INFN, Italy.


\end{document}